\documentclass[aps,prd,twocolumn,amsmath,amssymb,nofootinbib,preprintnumbers]{revtex4}

\usepackage[pdftex]{graphicx}
\usepackage{epstopdf}
\usepackage{amsmath}
\usepackage{amssymb}
\usepackage{subfigure}
\usepackage{hyperref}
\usepackage{url}
\usepackage{xcolor}
\usepackage{color}
\usepackage{graphicx}
\usepackage{gensymb}
\usepackage{tabularx}

\DeclareMathOperator{\sinc}{sinc}

\begin{document}
	
	\title{\textbf{Reflectivity and Spectrum of Relativistic Flying Plasma Mirrors}}
	
	\author{Yung-Kun Liu$^{1}$\footnote{d09222009@ntu.edu.tw},
		Pisin Chen$^{1,2}$\footnote{pisinchen@phys.ntu.edu.tw} and
		Yuan Fang$^{1}$\footnote{Current affiliation: Shanghai Jiao Tong University}  		
	}
	
	\affiliation{%
		~\\
		$^{1}$Department of Physics, National Taiwan University, Taipei 10617, Taiwan\\
		$^{2}$Kavli Institute for Particle Astrophysics and Cosmology, SLAC National Accelerator Laboratory, Stanford University, CA 94305, U.S.A.
	}%
	
\begin{abstract}
	Flying plasma mirrors induced by intense lasers has been proposed as a promising way to generate few-cycle EUV or X-ray lasers. In addition, if such a relativistic plasma mirror can accelerate, then it would serve as an analog black hole to investigate the information loss paradox associated with the black hole Hawking evaporation. Among these applications, the reflectivity, which is usually frequency-dependent, would affect the outgoing photon spectrum and therefore impact on the analysis of the physics under investigation. In this paper, these two issues are investigated analytically and numerically with one-dimensional particle-in-cell (PIC) simulations. Based on our simulation results, we propose a new model that provides a better estimate of the reflectivity than those studied previously. Besides, we found that the peak frequency of the reflected spectrum of a gaussian incident wave deviates from the expected value, $4\gamma^2\omega$, due to the dependence of reflectivity on the frequency of the incident wave. 
\end{abstract}

\maketitle

\newpage


\section{Introduction}

In 1905, Einstein studied \cite{Eistein1905} the interaction between light and a reflecting mirror moving with arbitrary velocity in vacuum. Since then, the concept of flying mirrors has attracted wide attention for theoretical and experimental applications. An optical laser pulse reflected from such a relativistic flying plasma mirror would not only upshift its frequency by a factor $4\gamma^2$, where $\gamma$ is the Lorentz factor of the mirror, into the x-ray regime, but also reduce its diffraction-limited volume due to the much shorter reflected wavelength. 

Relativistic flying mirror can be generated from irradiating an intense laser pulse on a plasma target. Such a plasma mirror is composed of a dense electron thin shell that moves with relativistic velocity. There are different mechanisms proposed to generate relativistic flying mirrors. Among them the idea of using the plasma wakefield \cite{Tajima:1979,Chen:1985} in the nonlinear perturbation regime as a relativistic flying plasma mirror \cite{BulanovSchwingerLimit} is particularly attractive. This scheme had been experimentally proven to be feasible by Kando et al. \cite{Kando:2007zz,Kando:2009,Pirozhkov2007}. Other ideas include, e.g. double-sided mirror \cite{doubleSidedMa}, oscillating mirror \cite{vincentiOscillatory}, sliding mirror \cite{slidingMirror}, nonlinear Langmuir waves \cite{Koga_2018,attoSecond}, and electron density singularities \cite{mu2019}).

The relativistic flying mirror has wide practical applications, ranging from higher harmonic generation (HHG) \cite{Koga_2018}, attosecond pulse generation \cite{attoSecond}, to XUV laser production \cite{Kando:2007zz,Kando:2009}. On the pure physics side, flying mirror with different trajectories has been theoretically investigated as a model to mimic phenomena associated with the quantum field theory in curved spacetime \cite{DaviesFulling1976,DaviesFulling1977,BirrellDavies}, which is closely associated with the physics of the black hole Hawking radiation \cite{Hawking1974}. On the basis of this theoretical analogy, an experimental scheme was recently proposed by Chen and Mourou \cite{ChenMourou:2017} with the intent to investigate Hawking evaporation through laser-driven relativistic flying plasma mirrors. In the proposal, an underdense plasma target with a tailored density gradient is invoked to realize a desired trajectory of the flying plasma mirror \cite{ChenMourou:2020}. Currently, there is an on-going project, AnaBHEL (Analog Black Hole Evaporation via Lasers),  that attempts to carry out such an experiment. If realized, it may help to shed more lights on the solution to the long-standing information loss paradox \cite{Hawking:1976}. 

Considering the reality that flying plasma mirrors tend to have a low reflectivity \cite{2013BulanovReview} and with a finite size, which deviate from the highly idealized theoretical studies in the literature, the analog Hawking radiation spectrum from flying plasma mirrors in 1+3D with a partial reflectivity have been calculated recently \cite{Kuan-Nan:2020, Chih-En:2020}, which should help to guide the design of the AnaBHEL experiment. To measure the trajectory of a flying mirror, which is closely related to the characteristic temperature of the Hawking radiation, the velocity of the plasma mirror at different instants (therefore different locations) should be measured so as to cross compared with the detected analog Hawking radiation spectrum.

To reconstruct the flying mirror trajectory and its relation to the reflected Hawking spectrum, the reflectivity and the Lorentz factor of the plasma mirror should be carefully studied. Martins et al. had discussed the dependence of the reflectivity on the electron density distribution of the plasma mirror \cite{Martins:2004}. Bulanov et al. have analytically investigated the reflectivity of a near-wave-breaking flying plasma mirror using the collisionless cold plasma theory \cite{2013BulanovReview}. From our one-dimensional particle-in-cell (1D PIC) simulations, we found that the reflectivity of the flying mirror is smaller than what Bulanov et al. predicted. This is due to the fact that the reflectivity depends sensitively on the electron density distribution of the plasma mirror, while that invoked in the theoretical model by Bulanov et al. differs from that in our PIC simulations. We therefore propose a fitting model based on the PIC data. This model can provide better prediction about the reflectivity of the flying plasma mirror. 

Furthermore, we extend the previous study on the interaction between a flying mirror and a plane wave to an incident wave with a finite bandwidth, which is closer to a realistic experimental setup. The peak frequency of the reflected spectrum from an incident wave with a Gaussian temporal profile is found to be deviated from the standard value of $[(1+\beta)/(1-\beta)]\omega_s$ due to the dependence of the reflectivity on the incident wave (source) frequency $\omega_s$. We suggest that the deviation can be treated as a correction term, which may serve to improve the precision of the Lorentz factor derived from the reflection spectrum. 
 
This article is organized as follows: in Section II we review previous studies on the reflectivity and propose a new model based on the numerical fitting of our PIC simulation data, which is different from that based on the near-wave-breaking condition. The validity of different density models are examined by analyzing the reflectivity of the flying mirror through 1D PIC simulations. Furthermore, we briefly describe the feasibility of generating water window X-ray from a relativistic flying mirror in underdense plasma. In Section III we discuss the reflected spectrum of the incident wave with Gaussian temporal profile and compare theoretical calculations with 1D PIC simulation results. Conclusions are given in Section IV.

\section{Reflectivity of a Flying Plasma Mirror}
The reflectivity of a relativistic flying plasma mirror has been studied by Bulanov \cite{2013BulanovReview}, Martins \cite{Martins:2004} and  H.-C Wu \cite{wu2009reflectivity}. Their procedure can be briefly described as follows. Consider a z-polarized planar electromagnetic wave with the vector potential $A_z=A_0\exp[i(\omega_st-k_sx)]$, where $A_0$, $\omega_s$ and $k_s$ are the amplitude, angular frequency and wave number of the incident wave, respectively. The vector potential satisfies the wave equation
 
\begin{align}
\label{eq:wave equation_Az}
\Big(\frac{\partial^2}{\partial t^2}-c^2\frac{\partial^2}{\partial x^2}\Big)A_z+\omega_p^2A_z=0,
\end{align}
where $\omega_p^2(x)\equiv n(x)e^2\mu_0c^2/m_e\gamma_{ph}$ is the plasma oscillatory frequency associated with the electron density distribution of the flying plasma mirror $n(x)$ with relativistic gamma factor $\gamma_{ph}$. Specifically, since the reflectivity is small, $A_z$ can be obtained perturbatively by either considering the system as weak reflection from a potential barrier \cite{Berry:1982,2013BulanovReview} or using Green's function method\cite{wu2009reflectivity}. In the calculation, it is convenient to perform Lorentz transformation from the lab frame to the mirror's proper frame, where ``$'$" denotes quantities in the mirror's proper frame. The amplitude ratio between incident wave and the reflected one is 

\begin{align}
	\frac{A_{r0}'}{A'_0}=\frac{i}{2k_s}\int_{-\infty}^{\infty}dx'\frac{\omega_p'^2}{c^2}e^{-2ik_s'x'}.
\end{align}
After substituting the definition of $\omega_p$, the ratio between the incident and reflected electric fields is

\begin{align}
\label{eq:E_general}
\frac{|E_r'|}{|E_0'|}=\frac{\mu_0ce^2}{2m_e\omega_s'}\int_{-\infty}^{\infty}dx\, n'(x)e^{2ik_s'x}.
\end{align}
From Eq.(\ref{eq:E_general}), it's clear that the reflected electric field depends on the density distribution of the flying plasma mirror.

We now investigate three different density distributions: Slab \cite{wu2009reflectivity}, Cusp \cite{bulanov2001relativistic}, and Square-Root Lorentzian Distribution (SRLD), defined as
\begin{align}
\label{eq:density_slab}
&\text{Slab : } \frac{n_{slab}(x)}{n_0}=\frac{C_{slab}}{D}\frac{c}{\omega_p}[H(x+2D)-H(x)],  \\
\label{eq:density_cusp}
&\text{Cusp: } \frac{n_{cusp}(x)}{n_0}=2^{1/3}\gamma \Big(3x\frac{\omega_p}{\beta_{ph}c}\Big)^{-2/3}-   \notag\\
&\quad\quad\quad\quad\quad\quad\frac{1}{60}\big(\gamma^4+2\gamma^3+71\gamma^2-20\gamma-60\big),\\
\label{eq:density_srld}
&\text{SRLD: } \frac{n_{srld}(x)}{n_0}=\frac{C_{srld}}{L}\frac{c}{\omega_p}\sqrt{\frac{L^2}{x^2+L^2}},
\end{align}
where $n_0$ and $\omega_p$ are the unperturbed background plasma density and the angular frequency, separately. To compare the results from different density distributions, we unify the definition of mirror density in these three distributions. For Slab and SRLD, the peak density and thickness of the mirror are associated with wave-breaking limit of the background plasma, under which the flying mirror contains half of the total electrons within the volume encompassed by the nonlinear plasma wavelength. The normalization constants are defined as $C_{slab}\equiv\sqrt{1+a_0^2/2}$ and $C_{srld}\equiv\sqrt{1+a_0^2/2}/[\sinh^{-1}(\lambda_{NP}/4L)]$, respectively. $\lambda_{NP}$ is the nonlinear plasma wavelength defined as $\lambda_{NP}\approx(2\sqrt{1+a_0^2/2}/\pi)\lambda_p$ \cite{BEREZHIANI1990338}, where $\lambda_p$ is the canonical plasma wavelength and $a_0\equiv eE_0/m_e\omega_0c$ the normalized vector potential of a linearly polarized driver pulse.
  
\begin{figure}[htp]
	\centering
	\includegraphics[height=175pt]{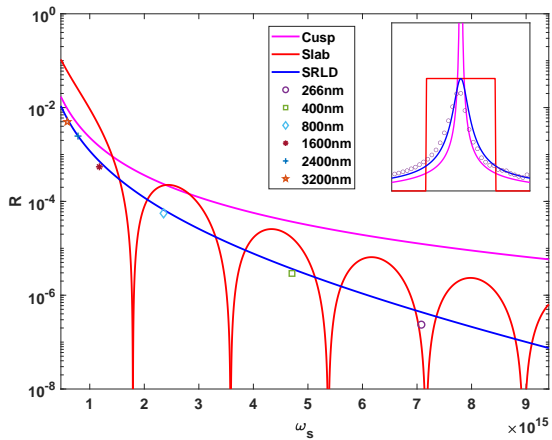}
	\caption{Reflectivity of the relativistic flying mirror as a function of the source frequency $\omega_s$. Solid lines are calculated from different reflectivity models (Eq.(\ref{eq:ref_slab})-(\ref{eq:ref_srld})). Distinct symbols are PIC simulation results with different $\lambda_s$. The SRLD model agrees well with PIC results and the Cusp model approaches SRLD in the long-wavelength limit of the source.   Inset: Comparison between three density distribution (Eq.(\ref{eq:density_slab})-(\ref{eq:density_srld})) models and the density of flying mirror from PIC simulations. Note that the PIC data (circles) agrees well with SRLD (blue line). }
	\label{fig:reflectivityAnaPIC}
\end{figure}

The respective parameters used and suitable scene of these different distributions are explained as below.
The Slab Distribution is a simplified model to describe the flying mirror, where $H(x)$ is the Heaviside step function and the thickness of the slab is defined as $2D$. The Cusp Distribution is derived from the 1D cold, collisionless plasma theory and the nonlinear coupled wave equation at the wave breaking situation \cite{2013BulanovReview}. $\gamma$ and $\beta$ are the Lorentz factor and the normalized velocity, respectively, calculated from the phase velocity of the flying mirror. The SRLD is a fitting function that we deduced from the PIC simulation. From 1D PIC simulations, the peak density of the flying mirror was found to be not as spiky as the Cusp Distribution but more rounded instead. Actually, the singularity in the Cusp Distribution at the wave-breaking point may suggest the breakdown of the cold plasma description. For a typical Laser Wakefield Accelerator scheme, warm plasma theory should take place as the flying mirror approaches the wave-breaking point, which in turn renders the maximum density finite \cite{SchroederWarmPlasma}. Without solving the complex equations based on the warm plasma theory, we deduced the SRLD distribution as a good approximation to the flying mirror density near the wave-breaking limit (see the Inset of Fig.(\ref{fig:reflectivityAnaPIC})). Here, $L$ is the characteristic thickness.

The reflectivity in terms of the photon number can be calculated from Eq.(\ref{eq:E_general}) and the result for the three different distributions are summarized as follows \cite{wu2009reflectivity,2013BulanovReview}
\begin{align}
\label{eq:ref_slab}
&\text{Slab: } R_{slab}(\omega_s)=\left[\frac{\omega_pC_{slab}}{4\gamma \omega_s}\sinc(4\gamma^2\omega_sD/c)\right]^2, \\
\label{eq:ref_cusp}
&\text{Cusp: } R_{cusp}(\omega_s)=\frac{\Gamma^2(2/3)}{2^2\cdot 3^{4/3}}\Big(\frac{\omega_{p}}{\omega_s}\Big)^{8/3}\frac{1}{\gamma^{4/3}},\\
\label{eq:ref_srld}
&\text{SRLD:} R_{srld}(\omega_s)=\left[\frac{\omega_pC_{srld}}{2\gamma\omega_s}K_0(4\gamma^2L\omega_s/c)\right]^2,
\end{align}
where the sinc function is defined as $\text{sinc}(x)\equiv\sin(x)/x$, $\Gamma$ is the gamma function and $K_0$ is modified Bessel function of the second kind  \cite{specialFunctionBook}. From Eq.(\ref{eq:ref_slab})-(\ref{eq:ref_srld}), it is clear that the reflectivity quickly decays as the frequency of incident wave $\omega_s$ increases. In addition, the reflectivity decreases as $\gamma$ increases. This means that there exists a trade-off between high reflectivity and high frequency in the reflected wave. 

The tendencies of such decrease in reflectivity are different among the three different density distributions of the sinc, the exponential ($\omega_s^{-8/3}$) and the $K_0$ functions, respectively. The decaying and oscillating behavior of the sinc function has been explained as the result of the modulations due to the constructive and destructive interferences \cite{wu2009reflectivity}. It should be noted that the argument in the sinc function and $K_0$ are of the same form, defined as $s\equiv 4\gamma^2L\omega_s/c=2\pi L/\lambda_r$, where $\lambda_r\equiv 2\pi c /(4\gamma^2\omega_s)$ is the reflected wavelength in the lab frame. As $s\gg1$, both sinc and $K_0$ functions decay quickly, which in turn highly suppress the reflectivity. Therefore, $s$ can serve as a parameter to define the quality of the flying mirror. A good mirror is one whose thickness is roughly the same order of magnitude as the doubly Doppler shifted wavelength, i.e., $L\leq O(\lambda_r)$. This explains why in an experiment one usually tunes the collision point at the wave-breaking limit so as to minimize the thickness of the flying mirror \cite{Kando:2007zz}\cite{Kando:2009}, which is an optimum point for trade-off between the reflectivity and the frequency of the reflected wave.

To examine the validity of Eq.(\ref{eq:ref_slab})-Eq.(\ref{eq:ref_srld}), we numerically study the property of relativistic flying mirror traversing a uniform plasma in the underdense regime with PIC simulations in 1D Cartesian geometry. The 1D configuration is a good approximation to the case of a driver pulse with a large focal spot in a higher dimension. This corresponds to the condition that $r\gg\lambda_p$ where $r$ is the spot radius of the driver pulse and $\lambda_p$ is the wavelength of the background plasma. The simulations are performed with the fully relativistic electromagnetic PIC code EPOCH \cite{epoch}.

In our simulation, the relativistic flying mirror is generated by a highly intense driver pulse (referred to as the "driver" from here on), which enters from the left boundary and propagates in the $+x$ direction. Along its way, the driver induces a flying mirror (wakefield) that follows behind it. The incident wave (referred to as the "source") enters, on the other hand, from the right boundary and propagates in the $-x$ direction. The collision point between the flying mirror and the source is tuned in such a way that the wave-breaking condition is reached with the flying mirror thickness minimized. Below we use subscripts ``m",``d",``s" ``r" to denote quantities that are associated with the flying mirror, the driver, the source and the reflected pulse, respectively.

The driver is characterized by the wavelength $\lambda_d=800{\rm nm}$ and the normalized vector potential $a_d=5.0$. The temporal profile is Gaussian with full-width-at-half-maximum (FWHM) duration of $\tau_d\approx\lambda_p/2$, which is chosen to excite the wakefield resonantly. The driver is linearly polarized with the electric field pointing in the y-direction.

To study the dependence of reflectivity on the source frequency, $\omega_s$. Planar wave source with several wavelengths are chosen: $266{\rm nm}$, $400{\rm nm}$, $800{\rm nm}$, $1600{\rm nm}$, $2400{\rm nm}$ and $3200{\rm nm}$. The normalized vector potential $a_s=0.004$ is set to be small enough to prevent the recoil effect \cite{Koga_2018,valenta2019recoil}. To distinguish the reflected pulse from the driver, we set the source linearly polarized in z-direction.
 
The background plasma density is uniform with a density $n_{p}=0.025n_c$, where $n_c\equiv m_e\epsilon_0\omega_d^2/e^2$ is the critical plasma density with respect to the driver. The simulation box size is $80\mu{\rm m}$ in the $x$ direction with $160,000$ cells. For shorter $\lambda_s$, the finer the grid size so as to guarantee the spatial resolution is sufficient for tracking the blue-shifted reflected pulse. In our strictest case, the resolution of the Cartesian grid size is roughly $8.3$ cells per reflected wavelength $\lambda_r$, which is estimated by $\lambda_r\approx\lambda_s/4\gamma_m^2$. Outflow conditions are applied to each simulation boundary for both electromagnetic waves and quasi-particles.

\begin{figure}[htp]
	\centering
	\includegraphics[height=175pt]{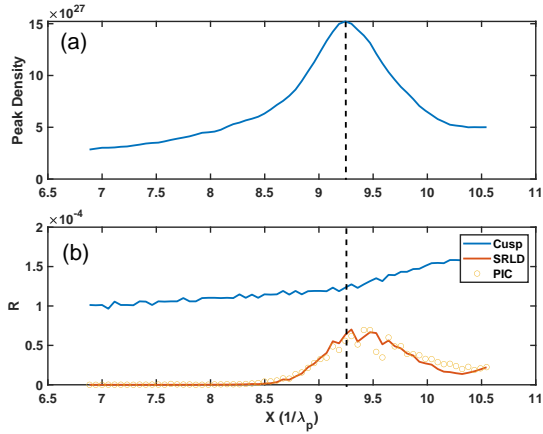}
	\caption{The evolution of parameters of the flying plasma mirror at different position. (a)The peak density  and (b) the instantaneous reflectivity of the flying mirror. In (b), the solid lines are the analytically predicted reflectivity based on Eq.(\ref{eq:ref_cusp}) and (\ref{eq:ref_srld}). Circles are the PIC simulation results. The black dashed line marks the position where wave breaks. }
	\label{fig:waveBreakingDefinition}
\end{figure}

It should be noted that a modified Cusp model was proposed in \cite{2013BulanovReview} and recovers Eq.(\ref{eq:density_cusp}) at the wave breaking point. Since the mirror is optimal at where wave breaks, we focus our discussion in this paper around that point. In the simulation, the wave breaking point is defined as the position where peak density of the flying plasma mirror reaches its maximum as Fig.(\ref{fig:waveBreakingDefinition}) shows. The following PIC results are all refered to the data at where wave breaks.

A comparison between the analytic formula and the simulation result on reflectivity is shown in Fig.(\ref{fig:reflectivityAnaPIC}). Parameters used in the analytic formulas (Eq.(\ref{eq:ref_slab})-(\ref{eq:ref_srld})) are $\omega_{p}=3.72\times10^{14} {\rm sec}^{-1}$, $a_d=5$, $n_0=4.35\times10^{25} {\rm m^{-3}}$, $\gamma=3.96$ and $L=0.96 {\rm nm}$. The first three parameters are fixed in the simulation setup while the last two are the values of the flying mirror at the wave breaking point.The rightmost PIC data point is the one with the source wavelength $\lambda_s=266 {\rm nm}$ which corresponds to the frequency tripling of the frequency of the conventional $800 {\rm nm}$ Ti:Sapphire laser. In this setup, the double-Doppler-shifted wavelength $\lambda_r\approx 4 {\rm nm}$ corresponds to the {\it water-window} X-ray wavelength, which can be a useful tool for life science research. The reflectivity in terms of the photon number is $R\approx2\times10^{-7}$, which is deduced from the ratio of the electric fields in the frequency spectrum between the reflected and the incident waves. 

We found that the reflectivity formula associated with the Cusp Distribution (Eq.(\ref{eq:ref_cusp})) approaches the simulation results in the long-wavelength limit of the sources ($\lambda_s \ge \lambda_d$) where the Cusp and the SRLD curves converge, as indicated by the leftmost point in Fig.(\ref{fig:reflectivityAnaPIC}) or the case in \cite{valenta2019recoil} with $\lambda_s=5\lambda_d$. However, as shown in Fig.(\ref{fig:reflectivityAnaPIC}), the discrepancy between Eq.(\ref{eq:ref_cusp}) and the PIC result grows as $\omega_s$ increases.
With $\lambda_s=266 {\rm nm}$ ($\lambda_s=\lambda_d/3$), we found a two orders of magnitude difference exists between Eq.(\ref{eq:ref_cusp}) and the PIC data. This is because the infinite electron density occurring at $x=0$ in Eq.(\ref{eq:density_cusp}) can in principle reflect higher frequency modes more effectively than those from finite peak densities such as that in Eq.(\ref{eq:density_slab}) and (\ref{eq:density_srld}). Within these three different models, the SRLD reflectivity formula gives the best agreement with the simulation results. This may not be surprising because one additional parameter, the thickness of the mirror, was introduced in SRLD.

In the simulation, the resolution of the cell depends on the reflected wavelength $\lambda_r$. For a higher source frequency, a higher resolution is needed. However, with the help of Eq.(\ref{eq:ref_srld}), one can estimate the reflectivity directly from the property of the flying mirror. This helps to greatly accelerate the process to search for an appropriate experimental parameter space.

\section{Reflected Spectrum and Deviation of Spectral Peak}

\begin{figure}[htp]
	\centering
	\includegraphics[height=150pt]{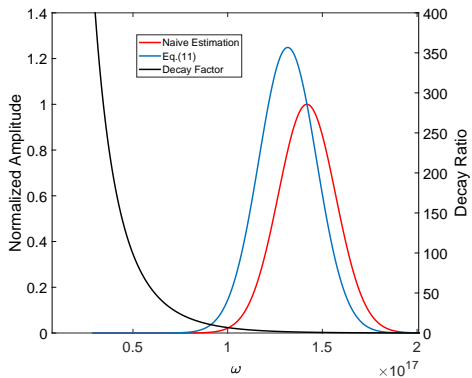}
	\caption{Normalized reflected electric field amplitude calculated by Eq.(\ref{eq:reflectedSpectrum}) (blue curve) and the naive estimation with $\omega=4\gamma_m^2\omega_s$ (red curve). The black curve shows the decaying term in Eq.(\ref{eq:reflectedSpectrum}) which is normalized to the value calculated with $\omega=4\gamma_m^2\omega_s$. The deviations of both frequency and amplitude at the peak of the spectrum are clearly seen.}
	\label{fig:deviationDemo}
\end{figure}

In the previous section, the incident wave was assumed to be a plane wave. In an actual experiment for XUV or attosecond pulse production, however, the incident laser pulses are expected to be short and therefore has a broad bandwidth. We shall model such a situation by a Gaussian wave packet with a central frequency $\omega_s$ and a pulse duration $\tau_s$:
\begin{align}
E_{in}(x,t)=E_0e^{-t^2/\tau_s^2}e^{i(k_sx-\omega_st)}.
\end{align}
With Eq.(\ref{eq:E_general}) and (\ref{eq:density_srld}), the electric field of the reflected wave from a flying mirror with a square-root Lorentzian distribution in the mirror's proper frame can be calculated. After transforming back to the lab frame, the reflected electric field is
\begin{align}
\label{eq:reflectedSpectrum}
\frac{E_r(\omega)}{E_0}=&2\sqrt{\pi}\tau_s L\gamma_mC_{crld}\frac{\omega_p}{\omega}K_0(L\omega/c) \notag\\
&\times\exp\Big\{\frac{-(\omega-4\gamma_m^2\omega_s)^2}{4}\Big(\frac{\tau_s}{4\gamma_m^2}\Big)^2\Big\}
\end{align}

where the ultra-relativistic limit is assumed, i.e., $(1+\beta_m)/(1-\beta_m)\rightarrow 4\gamma_m^2$, in order to simplify the equation. It should be noted that, when the background plasma density is sufficiently low ( $n_p/n_c \ll 1$), the parametric Doppler effect \cite{2013BulanovReview} due to the frequency dispersion in the background medium can be ignored. The exponential term describes a pulse with the central frequency at $4\gamma_m^2\omega_s$ and the pulse duration that is compressed by a factor $4\gamma_m^2$. However, the $\omega$-dependent and decaying term, i.e., $K_0(L\omega/c)/\omega$, will distort the reflected spectrum. Fig.(\ref{fig:deviationDemo}) shows the normalized reflected spectrum with $\gamma_m=5$, $L=15 {\rm nm}$, and $\lambda_s=800 {\rm nm}$. The red curve is the normalized exponential term. The blue curve is calculated from Eq.(\ref{eq:reflectedSpectrum}) and the black curve is the value of the decaying term. The distortion of the spectral shape, shown in the blue curve, is not evident, while both the frequency and the amplitude at the peak of the spectrum clearly deviate from the red curve. 

\begin{figure*}[htp]
	\centering
	\subfigure[Frequency deviation ratio as a function of the pulse duration]{
	\includegraphics[height=120pt]{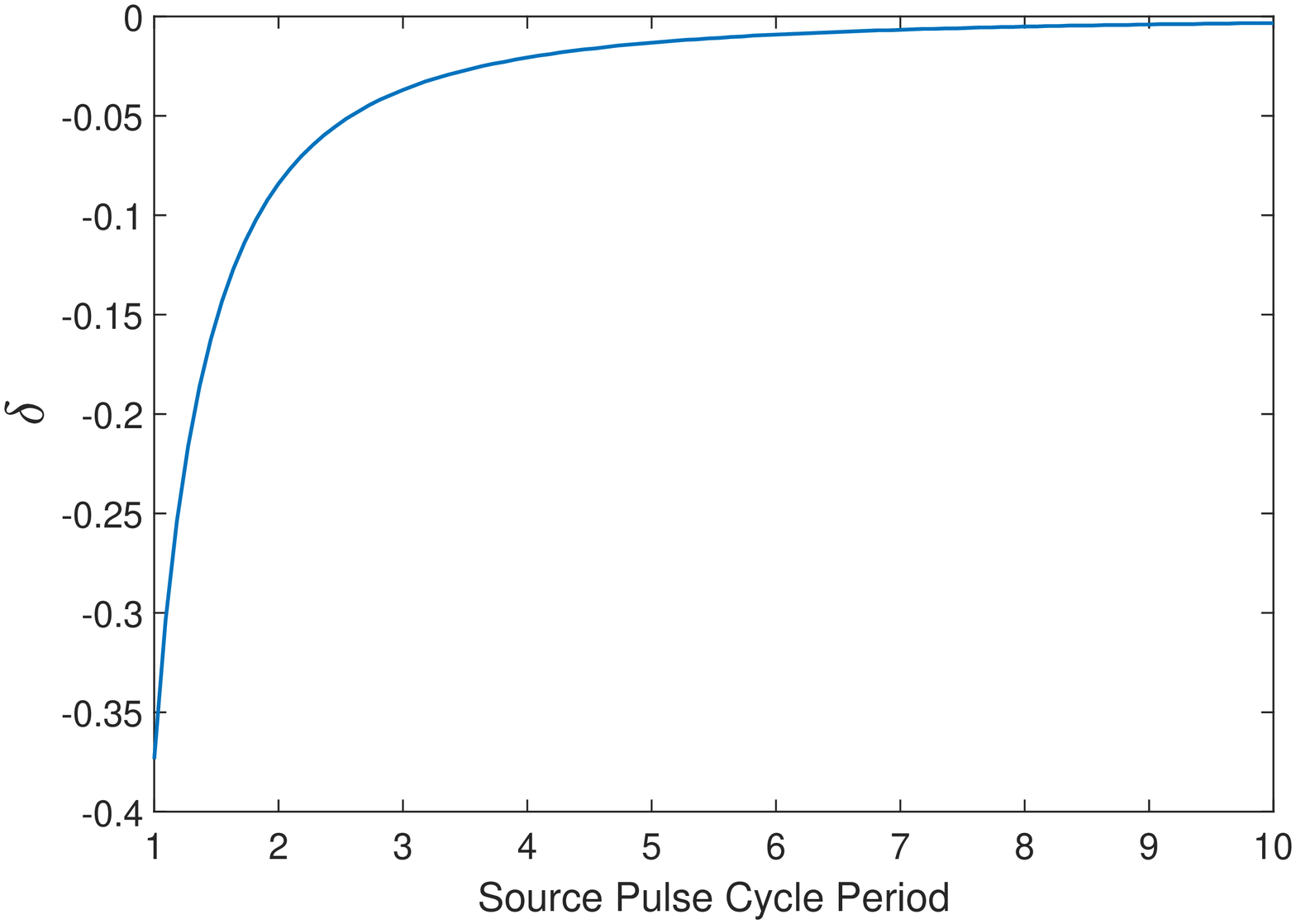}
	}
	\subfigure[Frequency deviation ratio as a function of $\gamma_m$]{
		\includegraphics[height=120pt]{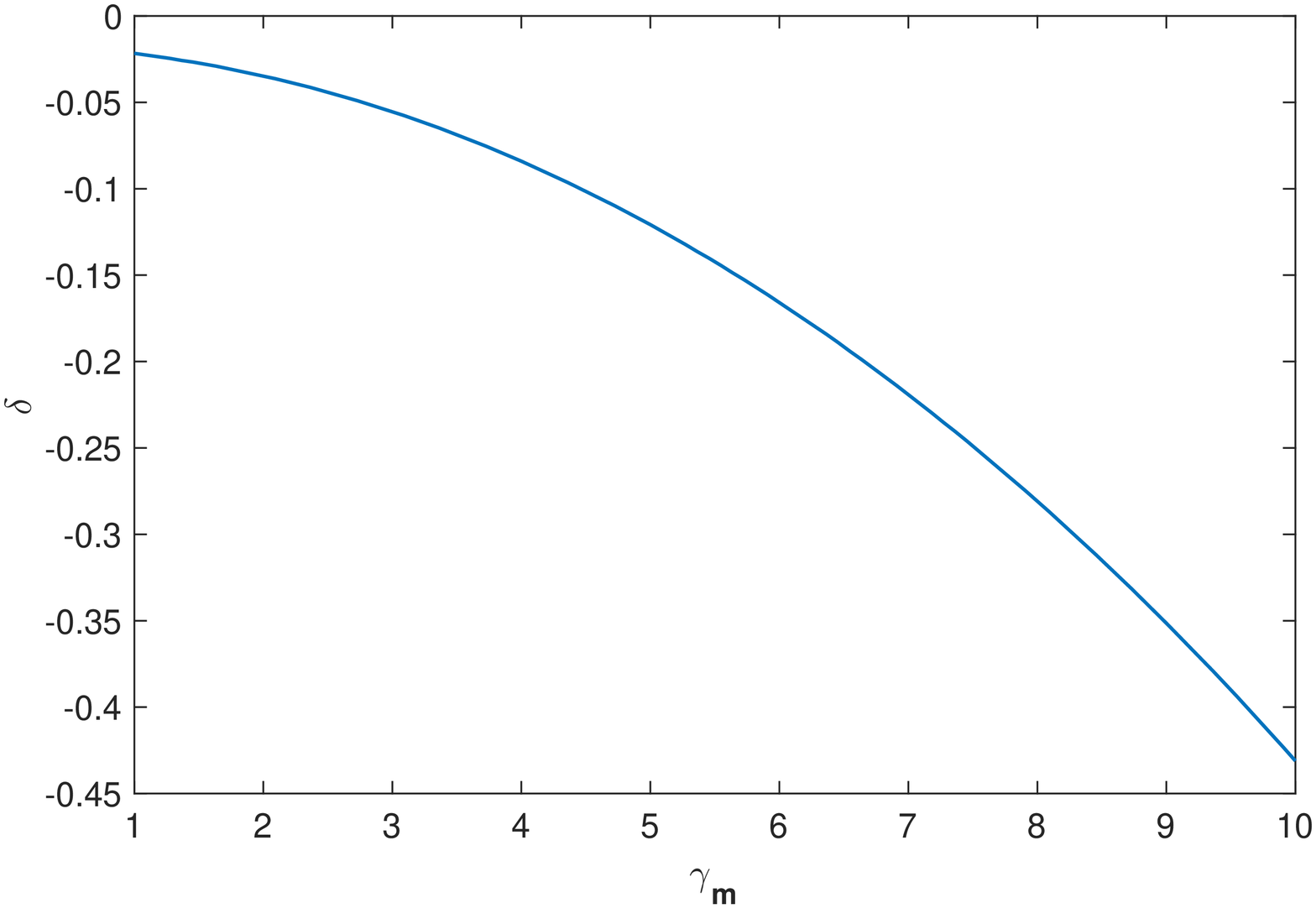}
	}
	\caption{Dependences of $\delta$ on the pulse duration $T$ and the Lorentz factor $\gamma$ with other parameters fixed. The deviation is evident for source pulses with few cycles or flying mirrors with high Lorentz factors.}
	\label{fig:dependence}
\end{figure*}

The deviation ratio between the frequency associated with the maximum amplitude, $\omega_{peak}$, and the naively estimated frequency, $\omega_{est}\equiv 4\gamma_m^2\omega_s$,  is defined as
\begin{align}
\delta\equiv\frac{\omega_{peak}-\omega_{est}}{\omega_{est}}
\end{align}
From Eq.(\ref{eq:reflectedSpectrum}), $\delta$ depends mainly on three parameters: the pulse duration of source $\tau_s$, the Lorentz factor of the flying mirror $\gamma_m$, and the characteristic thickness of the mirror $L_m$. Fig.(\ref{fig:dependence}) shows the dependence of $\delta$ on $\tau_s$ and $\gamma_m$, which are accessible in an experiment. $\tau_s$ can be measured with an auto-correlator and $\gamma_m$ can be estimated by the background plasma density, $\gamma_m\approx\omega_0/\omega_p$ \cite{Tajima:1979}, or the energy of the accelerated electrons \cite{Kando:2009,Pilloff}.

To have a sense about the amount of deviation, we make an estimation with Eq.(\ref{eq:reflectedSpectrum}) based on the typical parameters of intense lasers: $30 {\rm fs}$ pulse duration and $\gamma_m=4$. $\delta$ is found to be roughly $-1\%$, which may be hard to detect. However, from Fig.(\ref{fig:dependence}), it can be seen that the deviation is more significant for fewer-cycle sources and flying mirrors with higher Lorentz factors. This implies that the correction cannot be neglected when few-cycle pulses are employed or high blue-shift reflections are demanded, such as the situation for generating attosecond pulses with relativistic flying mirrors \cite{attoSecond}.

To study the validity of Eq.(\ref{eq:reflectedSpectrum}), the code EPOCH \cite{epoch} was used. Instead of flying mirrors induced by a driver laser, we imposed the mirror as an initial condition. The flying mirror was constructed as an electron sheet with a given longitudinal density distribution and propagating in the $+x$ direction with an initially assigned velocity. To prevent electrons from expelling each other during propagation, positive charge (proton) was introduced to co-move with the relativistic electron sheet. The interaction between protons and the source is negligible because of their large mass.  

We used the simplified Square-Root Lorentzian Distribution, $n_m(x)= n_{m,0}\sqrt{L_m^2/(x^2+L_m^2)}$, to characterize the density distribution of the flying mirror. There are three parameters to be determined: the peak density $n_{m,0}$, the characteristic thickness $L_m$, and the Lorentz factor of flying mirror $\gamma_m$. The peak density only affects the reflectivity. We therefore chose $n_{m,0}=3n_c$, where $n_c$ is the critical density for a $800 {\rm nm}$ electromagnetic wave, to guarantee that the reflected pulse is intense enough for observation. $\gamma_m=4$ and $L_m$ ranges from $1 {\rm nm}$ to $20 {\rm nm}$ according to the PIC results from the laser-driven flying mirror. The source is a linearly polarized pulse with a wavelength $\lambda_s=1.6 \mu{\rm m}$, which is long enough to increase the reflectivity. The normalized vector potential is $a_s=0.004$. The temporal profile is Gaussian with FWHM duration $\tau_s=1.66T_s$ where $T_s=\lambda_s/c$ is the source cycle period. The source enters from right boundary and propagates in $-x$ direction.The simulation box size is $50um$ in x direction with $50000$ cells. Therefore, the resolution of the Cartesian grid is $25$ cells per reflected wavelength, $\lambda_r\approx 4\gamma_m^2\lambda_s$. Boundary conditions remained the same as the setup in the previous section.

\begin{figure}[h]
	\centering
	\includegraphics[height=180pt]{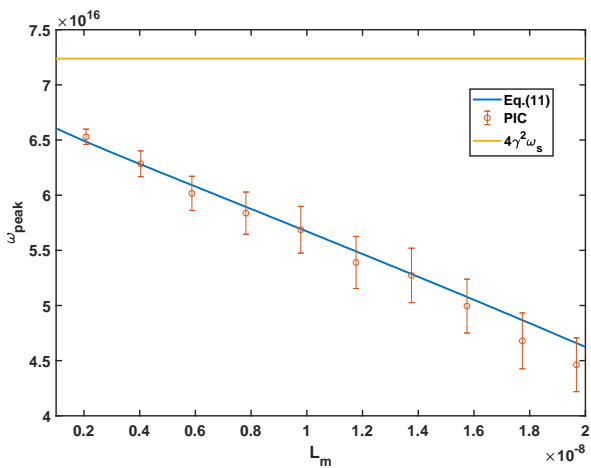}
	\caption{Comparison among the estimated double-Doppler shift frequency (yellow line), the theoretical prediction of the spectral peak $\omega_{peak}$ from Eq.(\ref{eq:reflectedSpectrum}) (blue line) and the PIC simulation results (red dots). The error bars of PIC results represent the uncertainty of statistical fluctuations in the initialization of the density distribution. $\gamma_m=4$, $n_{m,0}=3n_c$ and $\tau_s=1.66T_s$ are used as the initial condition. The 1D PIC simulation results agree reasonably well with the theoretical prediction. The linear dependence of the deviation on mirror thickness $L_m$ is also illustrated. }
	\label{fig:comparison}
\end{figure}

The comparison between the theoretical prediction of the frequency at the peak amplitude from Eq.(\ref{eq:reflectedSpectrum}) and the PIC simulation results is shown in Fig.(\ref{fig:comparison}). The horizontal yellow line is the estimated naive frequency $4\gamma_m^2\omega_s$ and the blue one is the maximum value of Eq.(\ref{eq:reflectedSpectrum}) solved numerically. Red circles with error bars are the PIC simulation results with different characteristic thicknesses $L_m$. The error bars represent the statistical fluctuations in the initialization of the SRLD distribution due to the limited number of macro-particles in our PIC simulations. We see that the PIC results are in reasonable agreement with the theoretical prediction of Eq.(\ref{eq:reflectedSpectrum}). From Fig.(\ref{fig:comparison}), the magnitude of the deviation, which is always negative, increases linearly as the characteristic thickness of the flying mirror $L_m$ increases, where the slope depends on the Lorentz factor, $\gamma_m$, and the source pulse duration, $\tau_s$. 

In an actual experiment, $\gamma_m$ can be deduced through the measurement of the reflected wave spectrum. Usually, this is estimated from the peak frequency of the reflected spectrum and the naive double-Doppler shifted relation, $\omega_{peak}=4\gamma_m\omega_s$. The deviation from this idealized value, as we have shown, can serve as its correction that can further improve the precision of this method.

We have shown in Section II that the plasma mirror thickness is an important parameter that determines the reflectivity and the reflected spectrum. In actual experiments, multiple tools can be employed to diagnose the dynamics of the wakefield, i.e., the mirror, such as the relativistic electron bunch probe \cite{electronProbe} and the optical probe \cite{opticalProbe}. However, the spatial and the temporal resolutions of these methods are still not precise enough to measure the thickness of a flying mirror near the wave-breaking condition, which is typically of tens of nanometer scale. Our investigation shows  that the frequency deviation can serve as a diagnosis on the thickness. As Eq.(\ref{eq:reflectedSpectrum}) shows, the peak frequency of the reflected wave depends on $\omega_s$, $\tau_s$, $\gamma_m$, and $L_m$. Among them $\omega_s$ and $\tau_s$ are laser parameters that can be measured accurately. In principle, $\gamma_m$ can be determined by conventional methods such as that based on the background plasma density \cite{Tajima:1979} or the accelerated electron energy \cite{Kando:2009,Pilloff}, from which the mirror thickness can be deduced. However, the diagnostic scheme suggested here may require highly stable condition of lasers and plasmas.

\section{Conclusion}
 
In this article, we extended previous studies on the reflectivity of relativistic flying mirrors with incident plane waves. We showed that the Square Root Lorentzian Distribution can accurately describe the flying mirror density distribution, and can provide a better estimation about the reflectivity. We defined a dimensionless parameter, $s= 2\pi L_m/\lambda_r$, to characterize the quality of the flying mirror. To attain a high enough reflectivity, the condition, $s\le2\pi$, must be satisfied, which means that the mirror must be thinner than the wavelength of the reflected pulse. In our simulations, we demonstrated the feasibility of  the generation of the water-window X-ray through plasma mirror reflection based on state-of-the-art laser parameters, which would provide great utility to life science researches. We found that the reflectivity in this case is $\sim2\times10^{-7}$ in photon numbers, which is encouraging. We also found that, for an incident wave with a Gaussian temporal profile, the peak frequency of the reflected spectrum is red-shifted from its expected value, $4\gamma_m^2\omega_s$. The magnitude of the deviation is positively correlated to the thickness of the mirror and its Lorentz factor, but negatively correlated with the duration of the source pulse. This deviation helps to provide a better description of the reflected spectrum, which can serve as a diagnostic tool about the dynamics of the wakefield. These studies about the reflectivity and the reflected spectrum may benefit future experiments such as AnaBHEL.

\section*{Acknowledgment}
This work is supported by ROC Ministry of Science and Technology (MOST) and Leung Center for Cosmology and Particle Astro-
physics (LeCosPA) of National Taiwan University. 


\end{document}